\documentclass[prl,twocolumn,showpacs,superscriptaddress,balancelastpage]{revtex4-1}
\usepackage[latin1]{inputenc}
\usepackage{graphicx}
\usepackage{amssymb}
\usepackage{amsmath}
\usepackage{xspace}
\usepackage{dcolumn}
\usepackage{bm}
\usepackage{color}
\usepackage[extra]{tipa}

\newfont{\tensy}{cmsy10}

\newcommand{\UP}[0]{\uparrow}
\newcommand{\DO}[0]{\downarrow}

\newcommand{\oP}{\hat{P}}
\newcommand{\oQ}{\hat{Q}}

\newcommand{\on}{\hat{n}}

\newcommand{\si}[0]{\sigma}
\newcommand{\om}[0]{\omega}

\newcommand{\kF}{k_\text{F}}
\newcommand{\kB}{k_\text{B}}
\newcommand{\nag}{{\phantom{\dag}}}

\newcommand{\psib}{\overline{\psi}}

\newcommand{\lcpi}{\lambda^\text{PI}_\text{c}}
\newcommand{\lcps}{\lambda^\text{PS}_\text{c}}

\newcommand{\las}[0]{\langle}
\newcommand{\ras}[0]{\rangle}

\newcommand{\la}[0]{\left\las}
\newcommand{\ra}[0]{\right\ras}
\newcommand{\ket}[1]{\left|#1\ra}
\newcommand{\bra}[1]{\la#1\right|}

\begin{document}


\title{%
Effect of Electron-Phonon Interaction Range for a Half-Filled Band in One Dimension}

\author{Martin Hohenadler}
\affiliation{%
\mbox{Institut f\"ur Theoretische Physik und Astrophysik, Universit\"at W\"urzburg,
97074 W\"urzburg, Germany}}

\author{Fakher F. Assaad}
\affiliation{%
\mbox{Institut f\"ur Theoretische Physik und Astrophysik, Universit\"at W\"urzburg,
97074 W\"urzburg, Germany}}

\author{Holger Fehske}
\affiliation{%
Institut f\"ur Physik, Ernst-Moritz-Arndt-Universit\"at Greifswald, 17487
Greifswald, Germany}

\begin{abstract}  
  We demonstrate that fermion-boson models with nonlocal interactions can be
  simulated at finite band filling with the continuous-time quantum Monte
  Carlo method. We apply this method to explore the influence of the
  electron-phonon interaction range for a half-filled band in one dimension,
  covering the full range from the Holstein to the Fr\"ohlich regime. The
  phase diagram  contains  metallic, Peierls, and phase-separated
  regions. Nonlocal interactions suppress the Peierls instability, and
  thereby lead to almost degenerate   power-law exponents for charge and
  pairing correlations.
\end{abstract} 

\date{\today}

 
\maketitle

{\it Introduction.}---Electron-phonon interaction has an essential influence on the properties
of many materials \cite{polaronbook2007}. It plays a key role for pairing and
superconductivity, mass renormalization, and charge ordering
phenomena. Taking into account quantum lattice fluctuations leads
to a complex, many-body problem. Consequently, theoretical studies usually
rely on simplified microscopic models.
A frequently invoked approximation, in particular for numerical
studies, is to consider a completely local electron-phonon coupling as in
Holstein's molecular-crystal model \cite{Ho59a}. However, nonlocal
interactions are expected to play an important role in materials with 
incomplete screening such as  quasi-one-dimensional (quasi-1D) organics \cite{Barford}. 
Long-range interactions, as described by the Fr\"ohlich model \cite{Fr54},
have been investigated in the context of high-temperature superconducting
cuprates \cite{Zehyer.90,0953-8984-14-21-308,Al.Ko.02,0034-4885-72-6-066501,Ha.Ha.Sa.Al.09},
and were found to support light polarons and bipolarons \cite{FeLoWe00,BoKaTr00,hague:037002}.

Exact numerical methods have played an important role for the understanding of
coupled electron-phonon systems. Whereas efficient algorithms exist for
Holstein-type models at arbitrary band filling
\cite{BlScSu81,HiFr82,0295-5075-84-5-57001,FeWeHaWeBi03,ClHa05,assaad:035116},
extended interactions could so far be addressed only in the
empty-band limit, see, for example, \cite{Ko98,FeLoWe00,PhysRevB.64.094507,MiPrSaSv00,MiNaPrSaSv03},
and \cite{0034-4885-72-6-066501} for a review. Consequently, key
phenomena such as the Peierls instability \cite{Peierls} were out of reach. 
The latter occurs in quasi-1D  systems with
commensurate fillings, for example in TTF-TCNQ \cite{doi:10.1021/cr030652g}, and drives a
transition to a Peierls insulator with charge-density-wave order. For the
Holstein model, it is known that quantum lattice fluctuations can destroy the
charge order \cite{JeZhWh99}, leading to a  metal-insulator quantum phase transition at
a finite value of the electron-phonon coupling strength. 

Unbiased investigations of the effect of nonlocal interactions at finite
band-filling represent a long-standing, open problem.  In this Letter, we use
the continuous-time quantum Monte Carlo (CTQMC) method \cite{Ru.Sa.Li.05} to study a model
that interpolates between and includes the paradigmatic Holstein and Fr\"ohlich limits.

{\it Model.}---We consider a Hamiltonian  $\hat{H} = \hat{H}_0 + \hat{H}_1$, where
$\hat{H}_0 = -t \sum_{\las ij\ras\sigma} (c^\dag_{i\si}c^\nag_{j\si} + \text{H.c.})$
describes 1D fermions with nearest-neighbor hopping $t$, and 
\begin{equation}\label{eq:modelH1}
  \hat{H}_1
  = 
  \sum_i \Big(\mbox{$\frac{1}{2M}$}\oP_i^2 + \mbox{$\frac{K}{2}$}\oQ_i^2\Big)
  -\gamma\sum_{i,r}f(r)\oQ_{i+r} \left(\on_{i}-1\right)
  .
\end{equation}
The first term describes lattice fluctuations in the harmonic
approximation, with the phonon frequency $\omega_0$ and the stiffness
constant $K=\omega_0^2 M$.
The second term represents the electron-phonon interaction, in the form of a
nonlocal density-displacement coupling, with the density operator
$\on_i=\sum_\si \on_{i\si}$ and $\on_{i\si}=c^\dag_{i\si} c^\nag_{i\si}$. The
matrix elements are chosen as \cite{AlKo99,FeLoWe00,Ha.Ha.Sa.Al.09} 
\begin{equation}
  f(r) = \frac{1}{(r^2+1)^{3/2}} \,e^{-r/\xi}
  \,,\quad
  0\leq r< L/2
  \,,
\end{equation}
where the lattice constant $a=1$.
For $\xi\to\infty$, this coupling represents a lattice version of the
Fr\"ohlich interaction \cite{AlKo99}. More generally, $\hat{H}_1$ may be
viewed as an extended Holstein interaction with screening length $\xi$;  the original
Holstein model \cite{Ho59a} is recovered in the limit $\xi\to0$. For
$\om_0\to\infty$, the model maps onto an attractive, generalized Hubbard model.
Our method can be applied to any coupling which preserves translational
invariance. The restriction of $r$ is due to periodic boundary conditions. 

{\it Method.}---For
electron-phonon problems \cite{assaad:035116}, the starting point is the
partition function at inverse temperature $\beta=1/\kB T$,
\begin{equation}
Z = \int \mathcal{D}(\psib,\psi)
e^{-S_0[\psib,\psi]}
\int\mathcal{D}(q)
e^{- S_{1}[\psib,\psi,q]}\,,
\end{equation}
where $\psib,\psi$ are Grassmann fields, and $q$ denotes phonon coordinates.
The phonons can be integrated out {\it exactly} \cite{Feynman55}, leading to
a purely fermionic action with a nonlocal (in space and time) interaction
\begin{equation}\label{eq:action}
S^\text{f}_1= - \iint_0^\beta d\tau d\tau' 
\sum_{ij}\, [n_i(\tau)-1]
D_{i,j}^{\tau,\tau'}
[n_j(\tau')-1]\,.
\end{equation}
The phonon propagator takes the form
$D_{i,j}^{\tau,\tau'}=F(i-j) D(\tau-\tau')$ with $F(i-j)= \sum_{r} f(r+i-j)
f(r)$ and the Holstein propagator $D(\tau-\tau')$. The interaction
range in space (time) is determined by $\xi$ ($\omega_0$). The CTQMC method used here
is based on an exact expansion around $\gamma=0$
\cite{Ru.Sa.Li.05}. A hybridization expansion algorithm for
electron-phonon impurity problems also exists \cite{werner:146404}.
Monte Carlo updates consist of adding or removing single vertices,
and flipping auxiliary Ising spins \cite{Ru.Sa.Li.05,assaad:035116}.
The numerical effort scales with the cube of the average expansion order.
Because of the underlying weak-coupling expansion, the CTQMC method is
particularly efficient for problems with small
expansion orders, such as the Peierls transition in the adiabatic
regime $\om_0/t\ll 1$ \cite{HoFeAs11}.  Importantly, the method enables us to
study the many-electron problem \cite{HoNevdLWeLoFe04}. We have verified that
it reproduces exact results in the Holstein limit $\xi\to0$, and in the
anti-adiabatic limit $\om_0\to\infty$ where the model~(\ref{eq:modelH1}) maps
onto an extended attractive Hubbard model.

{\it Results.}---We choose a phonon frequency $\om_0/t=0.5$. In this regime,
representative of many materials, neither static nor instantaneous
approximations are valid, and numerical simulations are essential.
The dimensionless ratio $\lambda=\epsilon_\text{p}/2t$, where
$\epsilon_\text{p}$ is the polaron binding
energy in the atomic limit ($t=0$) \cite{AlKo99} and $2t$ is half of the
free bandwidth,  allows us to compare different $\xi$ at the same
effective coupling strength. We use $t$ as the energy unit, and set
$\hbar=M=1$. All results are for a half-filled band. 

\begin{figure}[t]
  \includegraphics[width=0.425\textwidth,clip]{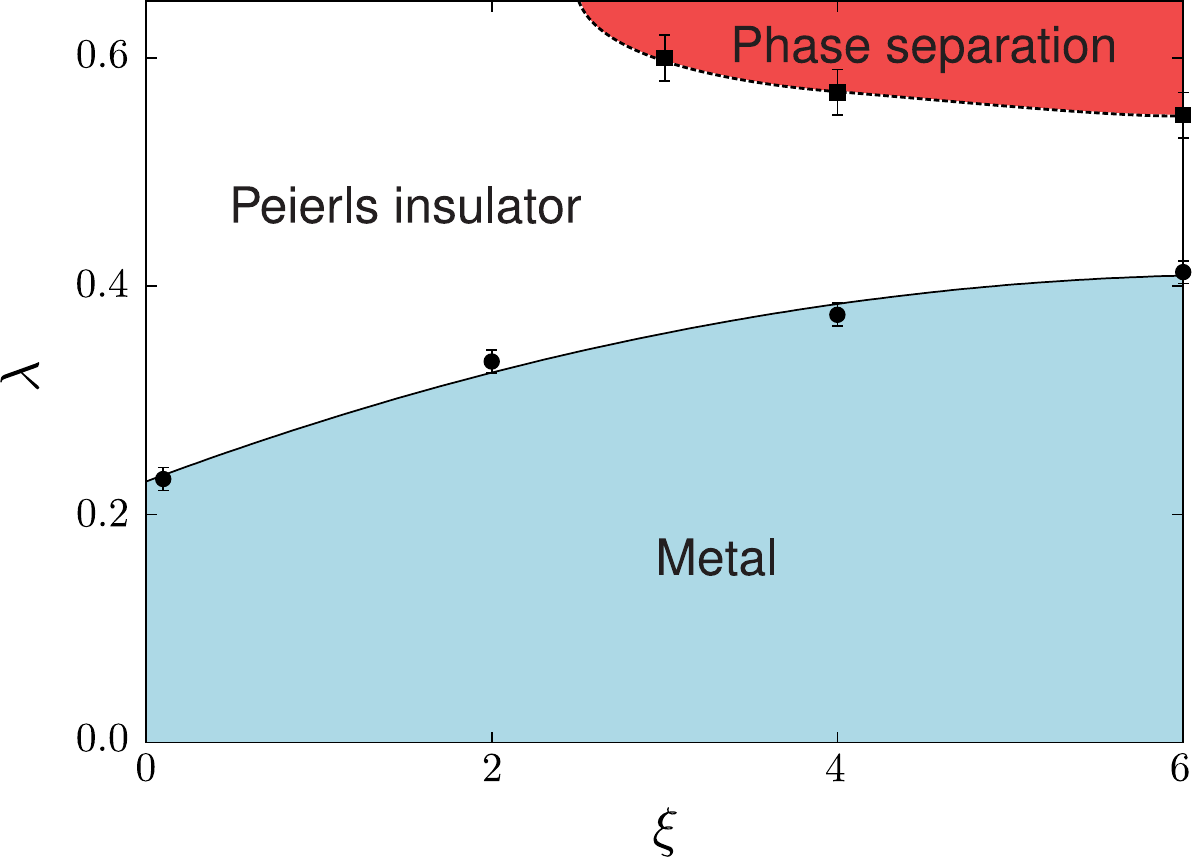}
  \caption{\label{fig:pd}
    (Color online) Phase diagram as a function of interaction range $\xi$ and
    electron-phonon coupling strength $\lambda$. The regions correspond to
    a metal, a Peierls insulator, and phase separation. The metallic and
    Peierls phases (and presumably also phase separation) extend to the
    Fr\"ohlich limit $\xi=\infty$. Lines are guides to the eye. Here 
    $\om_0/t=0.5$.
  }
\end{figure}

\begin{figure}[t]
  \includegraphics[width=0.45\textwidth]{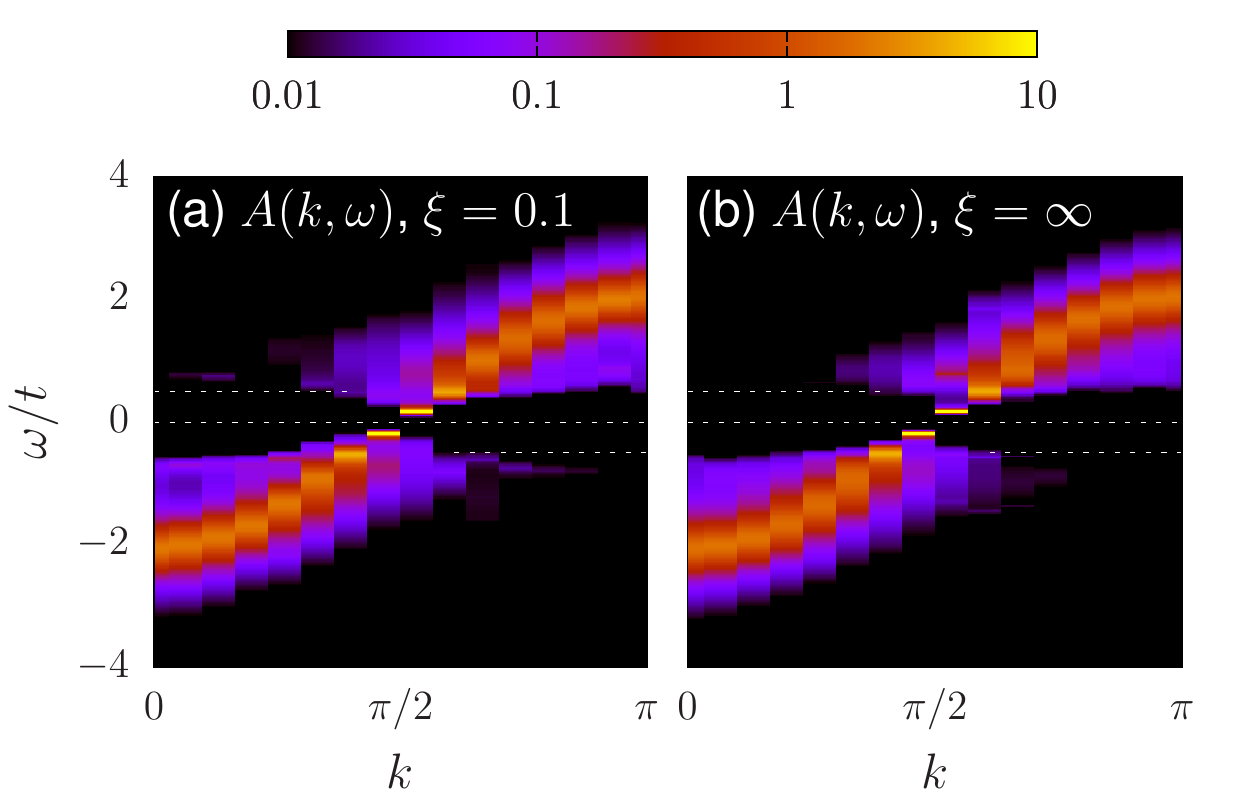}
  \caption{\label{fig:akw_LL}
    (Color online) Single-particle spectral function $A(k,\om)$
    [Eq.~(\ref{eq:akw})] in the metallic phase at $\lambda=0.2$ for (a)
    $\xi=0.1$, (b) $\xi=\infty$. Here $\om_0/t=0.5$, $\beta t=L=30$. The
    dashed lines indicate the Fermi level ($\om=0$) and $\om=\pm \om_0$.
  }
\end{figure}

The phase diagram as a function of $\xi$ and $\lambda$, obtained from CTQMC
simulations with up to $L=42$ sites, is shown in Fig.~\ref{fig:pd}. Because
the Holstein model is recovered for $\xi=0$, its previously studied metallic
and Peierls insulating phases
\cite{0295-5075-84-5-57001,hardikar:245103,PhysRevB.76.155114}
smoothly extend to $\xi>0$. However, we observe a significant $\xi$-dependence for
small $\xi$, and saturation for larger values. Additionally, for sufficiently
large $\xi$ and $\lambda$, we find a region of phase separation or charge
segregation. The metallic phase and the metal-insulator transition extend all
the way to the Fr\"ohlich limit $\xi=\infty$ [$\lcpi=0.48(2)$]; see also
Fig.~\ref{fig:akw_LL}. For the range of $\lambda$ shown in
Fig.~\ref{fig:pd}, phase separation is absent for $\xi\leq2$, and becomes
more favorable with increasing $\xi$.

\begin{figure*}
  \includegraphics[width=0.85\textwidth]{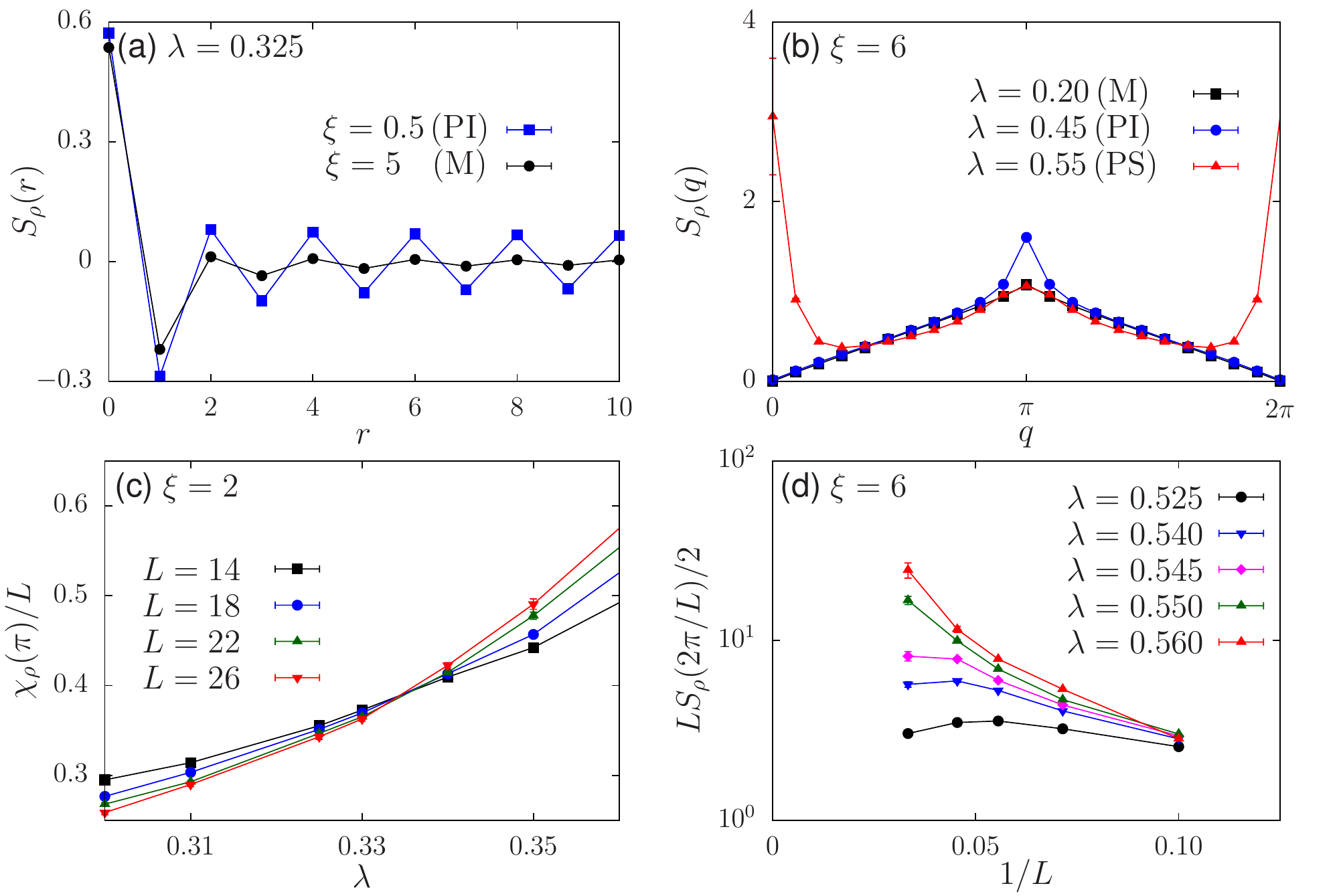}
  \caption{\label{fig:details}
    (Color online) (a) Density-density correlation function $S_\rho(r)$, for
    $\lambda=0.325$ and $\beta t=L=22$, in the Peierls insulator (PI, $\xi=0.5$) and the metallic
    phase (M, $\xi=5$). (b) Structure factor $S_\rho(q)$ at $\xi=6$ in the three
    phases of Fig.~\ref{fig:pd}. We used $L=22$, and $\beta=L$ ($L/2$) for
    $\lambda=0.2$, 0.45 (0.55). (c)
    Scaling of the charge susceptibility $\chi_\rho(\pi)$
    [Eq.~(\ref{eq:chic})] at $\xi=2$, defining the critical point
    $\lcpi=0.33(1)$ of the Peierls transition. Here $\beta=L$. (d)
    Finite-size scaling of $\pi S_\rho(q_1)/q_1=LS_\rho(2\pi/L)/2$ 
    with $\beta=L/2$. The divergence for $\lambda\geq 0.55$
    indicates phase separation (PS).  All results are for $\om_0/t=0.5$.
  }
\end{figure*}

The different phases can be characterized by the density correlator
$S_\rho(r)=\las (\on_r-1)(\on_0-1)\ras$ and the density structure factor
$S_\rho(q)$.  In the metallic phase [Fig.~\ref{fig:details}(a)],
$S_\rho(r)$ shows a power-law decay of $2\kF$ correlations (with exponent
$K_\rho$) as expected from bosonization, and is linear for $q\to0$. Together
with exponentially suppressed spin correlations (not shown), these findings
are consistent with a bipolaronic Luther-Emery phase \cite{Voit.94,1742-6596-200-1-012031}.
 
The Peierls state exhibits quasi-long-range $2\kF$ density correlations
[Fig.~\ref{fig:details}(a)],  corresponding to charge-density-wave order at
$T=0$ with two electrons of opposite spin forming bipolarons on
every other site. The phase boundary for the Peierls transition can be determined from the
staggered charge susceptibility \cite{hardikar:245103} 
\begin{equation}\label{eq:chic}
  \chi_\rho(\pi) = \frac{1}{L} \sum_{ij} (-1)^{i-j} \int_0^\beta d\tau
  \las \on_i(\tau) \on_j(0) \ras\,.
\end{equation}
For fixed $\beta/L$, $\chi_\rho(\pi)/L$ is universal at the critical point, and
the crossing of curves for different $L$ gives $\lcpi$; for example, we have 
$\lcpi=0.33(1)$ for $\xi=2$ in Fig.~\ref{fig:details}(c). As for the
Holstein model \cite{HiFr83II,hardikar:245103}, the Peierls transition is
expected to be of the Kosterlitz-Thouless type also for $\xi>0$.
Extended interactions ($\xi>0$) promote metallic behavior by dissociating the
onsite bipolarons predominant in the Holstein regime, and we see a
Peierls insulator to metal transition as a function of $\xi$
[Fig.~\ref{fig:details}(a)]. By the same mechanism, the critical coupling for the
transition to the Peierls insulator, $\lcpi$, increases
with increasing $\xi$, see Fig.~\ref{fig:pd}.  The onset of charge order is
also reflected in the divergent $q=2\kF$ peak in $S_\rho(q)$, see Fig.~\ref{fig:details}(b).

Phase separation  as a result of the phonon-induced attraction manifests
itself as a peak at small $q$ in $S_\rho(q)$, as shown in Fig.~\ref{fig:details}(b).
In the phase-separated region of Fig.~\ref{fig:pd}, the quantity $\pi
S_{\rho}(q_1)/q_1$ (with $q_1 = 2\pi/L$)---whose thermodynamic limit is
related to $K_\rho$ in a Luttinger liquid---diverges with system size. 
In the Peierls phase $K_\rho=0$, as verified on very large systems
\cite{hardikar:245103}.  Formally, $K_\rho=\infty$, reflecting phase
separation, implies a divergent compressibility \cite{Voit.94}. The divergence of $\pi
S_{\rho}(q_1)/q_1$ in the phase-separated region is shown for $\xi=6$ in
Fig.~\ref{fig:details}(d), and we deduce a critical value of $\lcps=0.55(2)$.
There are two possible scenarios for the transition from the Peierls to the
phase separated region. A continuous transition would imply a melting of
the charge-ordered Peierls state before the formation of multipolaron
droplets, allowing for an intervening (narrow) metallic
region with finite $K_\rho$. Alternatively, the insulator-insulator transition could be of
first order. Evidence for the latter possibility comes from the occurrence of
metastable configurations and hysteresis at low temperatures in our simulations.
The occurrence of phase separation  in models with long-range electron-phonon
coupling had been suggested before \cite{PhysRevB.64.094507} and observed in
analytical work \cite{Al.Ko.02,0953-8984-14-21-308}; for short-range
interactions,  it is suppressed by the absence of bound triplet states
\cite{PhysRevB.64.094507}, but may occur in the vicinity of a Mott transition
\cite{PhysRevLett.92.106401}. Phase  separation is expected to be suppressed
in the presence of additional long-range electron-electron interaction
\cite{Al.Ka.2000}.

In combination with analytical continuation \cite{Be.04},
we can calculate the single-particle spectral function
\begin{equation}\label{eq:akw}
  A(k,\om)
  =
  \frac{1}{Z}\sum_{ij}
  {|\bra{i} c_{k\sigma} \ket{j}|}^2 (e^{-\beta E_i}+e^{-\beta E_j})
  \delta(\Delta_{ji}-\om)
  \,,
\end{equation}
where $\ket{i}$ is an eigenstate with energy $E_i$, and
$\Delta_{ji}=E_j-E_i$. Results in the metallic phase ($\lambda=0.2$) are shown in
Fig.~\ref{fig:akw_LL} for the extreme Holstein ($\xi=0.1$) and
Fr\"ohlich limits ($\xi=\infty$). The locus of spectral weight follows
the free band dispersion, $-2t\cos k$. The exponentially small Luther-Emery
spin gap is not resolved for the parameters chosen, and the spectrum is
particle-hole symmetric. Excitations
are sharp inside the coherent interval $[-\om_0,\om_0]$, whereas they are
substantially broadened as a result of multiphonon processes at higher
energies \cite{LoHoFe06}. As expected for our 1D model, the spectrum agrees
well with the exact bosonization result \cite{MeScGu94}, which
predicts a hybridization of the spin, charge and phonon modes, although
spin-charge separation is not visible due to the weak coupling and
small $\om_0$ \cite{NiZhWuLi05}. Comparing Figs.~\ref{fig:akw_LL}(a) and (b)
we see that in contrast to previous work on one and two electrons
\cite{Ko98,FeLoWe00,PhysRevB.64.094507,MiPrSaSv00,MiNaPrSaSv03,0034-4885-72-6-066501},
the impact of the interaction range is remarkably small. This important
characteristic of the many-electron case can be related to the absence of
significant polaron and bipolaron effects in the metallic phase of Fig.~\ref{fig:pd}. 

\begin{figure}[t]
  \includegraphics[width=0.45\textwidth,clip]{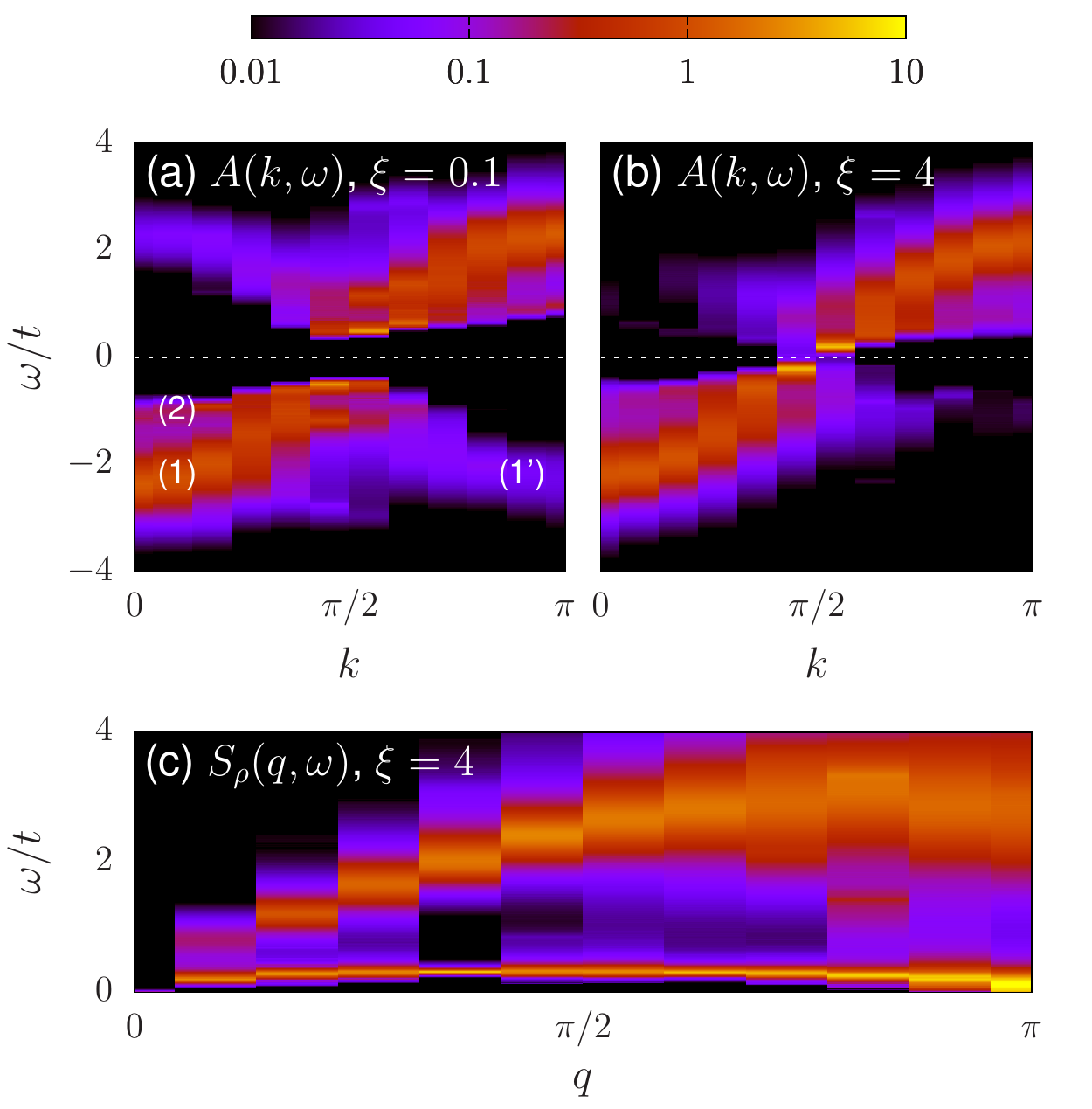}
  \caption{\label{fig:akw_PI}
    (Color online) (a),(b) Single-particle spectral function $A(k,\om)$ in
    the Peierls phase at $\lambda=0.4$ for (a) $\xi=0.1$, (b) $\xi=4$. (c)
    Dynamical charge structure factor [Eq.~(\ref{eq:nqw})] for the same
    parameters as in (b). Here $\om_0/t=0.5$, $\beta t=L=22$. The dashed
    lines indicate (a),(b) the Fermi level and (c) $\om_0$. The labels
    (1), (1'), and (2) in (a) are explained in the text. 
  }
\end{figure}

Figures~\ref{fig:akw_PI}(a),(b) show $A(k,\om)$ in the Peierls
phase ($\lambda=0.4$). In the Holstein regime ($\xi=0.1$), the spectrum
consists of two sets of features. The cosine band seen in
Fig.~\ref{fig:akw_LL} has acquired a gap at the Fermi level and reveals
additional, backfolded shadow bands as a result of dimerization and the
corresponding doubling of the unit cell \cite{Vo.Pe.Zw.Be.Ma.Gr.Ho.Gr.00}. These
signatures are labeled (1) and (1') in Fig.~\ref{fig:akw_PI}(a),
respectively. In addition, we find lower-energy excitations labeled (2)
corresponding to bound soliton-antisoliton pairs or, equivalently,
polarons \cite{RevModPhys.60.781,HoFeAs11}, which are absent in a
homogeneous mean-field solution that captures only (1) and (1') \cite{HoFeAs11}.
The soliton dispersion indicates a mass larger than the electron
mass. Because their energy at the Fermi level is lower than the Peierls gap,
doping would lead to the formation of solitons \cite{SuShHe79}.

Increasing the interaction range from $\xi=0.1$ to $\xi=4$ drives
the system into the vicinity of the metal-insulator transition, see
Fig.~\ref{fig:pd}. This is reflected by a much smaller gap at the Fermi
level, reduced spectral weight of the polaron excitations, and the
suppression of shadow bands. Consequently, the spectral function becomes
quite similar (but not identical) to that shown in Fig.~\ref{fig:akw_LL}, and
illustrates the continuous evolution of $A(k,\om)$ across $\lcpi$.

A hallmark feature of the Peierls state is phonon softening at $q=2\kF$,
which is visible \cite{HoFeAs11,assaad:155124} in the dynamical charge structure
factor  [$\hat{\rho}_q = \sum_r e^{iqr} (\on_{r} - \las \on_r\ras) /\sqrt{L}$]
\begin{equation}\label{eq:nqw}
  S_\rho(q,\om)
  =
  \frac{1}{Z}\sum_{ij} {|\bra{i} \hat{\rho}_q \ket{j}|}^2
  e^{-\beta E_j} 
  \delta(\Delta_{ji}-\om)
  \,.
\end{equation}
The results in Fig.~\ref{fig:akw_PI}(c) for $\lambda=0.4$, $\xi=4$ [the same
parameters as in Fig.~\ref{fig:akw_PI}(b)] reveal a clear signature of the
renormalized phonon dispersion. The spectrum is dominated by the soft phonon
mode at $\om=0$, $q=2\kF$. Furthermore, we observe a continuum of
particle-hole excitations, and a charge gap at long wavelengths ($q\to0$).

\begin{figure}[t]
  \includegraphics[width=0.45\textwidth,clip]{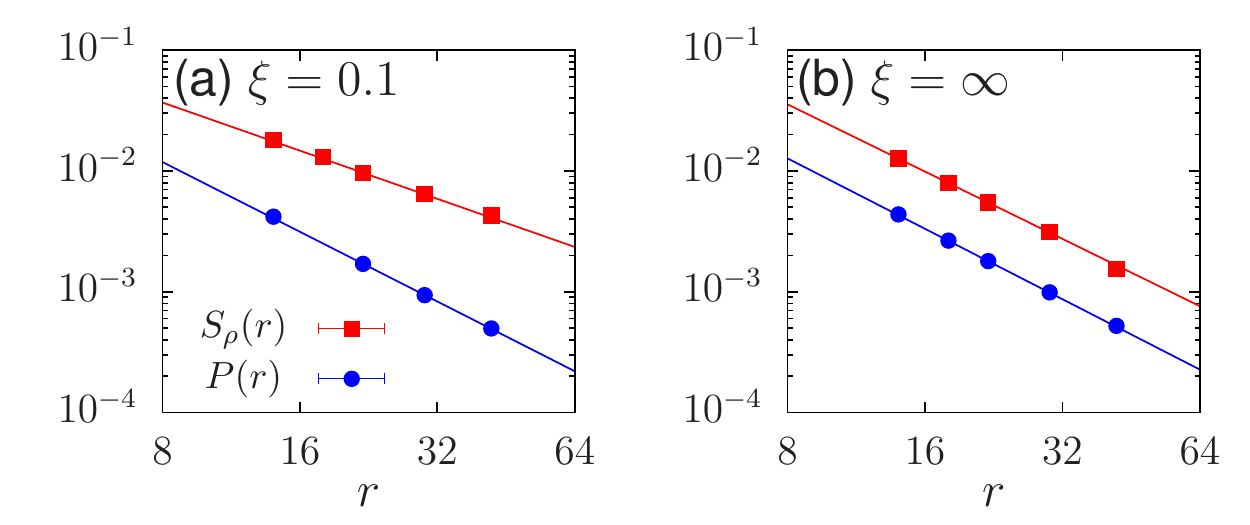}
  \caption{\label{fig:pairing}
    (Color online) Charge and pairing correlations at distance $r=L/2$
    for (a) $\xi=0.1$, (b) $\xi=\infty$. Here $\om_0/t=0.5$ and $\lambda=0.2$. Lines are fits
    to a power law $f(r)=A r^{-\eta}$.
  }
\end{figure}

Finally, we consider the interaction-range effect on the competition of
charge and pairing correlations. Figure~\ref{fig:pairing} shows results for
$S_\rho(r)$ and the s-wave pair correlator $P(r)=\las \Delta^\dag_r
\Delta^\nag_0\ras$ (with $\Delta^\dag_r= c^\dag_{r\UP} c^\dag_{r\DO}$)
in the metallic phase. For $\xi=0.1$, Fig.~\ref{fig:pairing}(a) reflects the dominance
of charge correlations previously observed for the Holstein model 
\cite{PhysRevB.84.165123}. However, Fig.~\ref{fig:pairing}(b) reveals
that long-range electron-phonon interaction (here $\xi=\infty$) suppresses
charge correlations and thereby results in almost identical power-law
exponents in both channels. Additional short-range electron-electron repulsion
is expected to further suppress onsite bipolaron formation---promoting
nonlocal pairing---and to lead to a more general phase diagram with Mott, metallic,
Peierls and phase-separated ground states. It will be interesting to explore
this issue further, both at and away from half filling, also in the light of
a recently reported phase of correlated singlets \cite{Re.Ya.Li.11}.

{\it Conclusions.}---We used exact Monte Carlo simulations to study
many-electron systems with nonlocal and even long-range electron-phonon
interaction. Compared to the Holstein model, extended interactions suppress
the Peierls instability---making pairing more favorable---and can lead to
phase separation. Key implications for materials modeling are that
interactions of finite but small range are well described by Holstein-type
models, whereas long-range interactions can have substantial effects on
the balance of pairing and charge correlations.

We are grateful to F. Essler and A. Muramatsu for helpful
discussions, and acknowledge support from the DFG Grant No.~Ho~4489/2-1
as well as computer time at the LRZ Munich and the J\"ulich
Supercomputing Centre.


\end{document}